\documentclass[12pt]{article}
\usepackage{latexsym}
\usepackage{graphicx}
\usepackage{epstopdf}
\usepackage{amsthm}
\title{Negative closed walks in signed graphs: A note}
\author{Andrzej Szepietowski \\
         Institute of Informatics,\\ 
 Faculty of Mathematics, Physics, and Informatics, \\
University of Gda\'nsk, 
80-308 Gda\'nsk, Poland \\
                {\tt matszp{@}inf.univ.gda.pl}}

\date{}
\begin{document}
\maketitle
\newtheorem{df}{Definition}
\newtheorem{lem}[df]{Lemma}
\newtheorem{thm}[df]{Theorem}
\newtheorem{cor}[df]{Corollary}

\begin{abstract}
Recently Naserasr, Sopena, and Zaslavsky 
[R. Naserasr, \'E. Sopena, T. Zaslavsky,
{\sl Homomorphisms of signed graphs: An update,}
arXiv: 1909.05982v1 [math.CO] 12 Sep 2019.]
published a report on closed walks in signed graphs. They gave  
a characterization of the sets of closed walks in a graph $G$ which corespond to the set of negative walks in some signed graph on $G$. In this note we show that their characterization is not valid and give a new characterization.
\end{abstract}
\section{Introduction}
Recently Naserasr, Sopena, and Zaslavsky \cite{NSZ} published a report on 
closed walks in signed graphs. A signed graph $(G,\sigma)$ is a graph  
$G=(V(G),E(G))$ (with loops and multiple edges allowed) and a signature function $\sigma: E(G)\to\{+,-\}$ which marks each edge with $+$ or $-$.
Let us consider the set ${\cal W}_\sigma$ of negative closed walks in $(G,\sigma)$, i.e. walks with odd number of edges marked with $-$.
It is easy to see that the set ${\cal W}_\sigma$ is closed under rotatian.
I.e. for every closed walk  $W\in{\cal W}_\sigma$ every rotation of $W$
is in ${\cal W}_\sigma$. A rotation of $W$ is a walk which has the same sequence of edges but it starts in a different vertex.
Moreover, the set ${\cal W}_\sigma$ satisfies so called exclusive 3-walk property. In order to present the property we shall need the following function:

\begin{df}
For a set $\cal W$ of closed walks in a graph $G$, the function $\sigma_{\cal W}$ is defined by:

$ \sigma_{\cal W}(W)= +$\quad if\,  $W\notin {\cal W}$,

$ \sigma_{\cal W}(W)= -$\quad  if\,  $W\in {\cal W}$.
\end{df}

\begin{df}[Exclusive 3-walk property]
A set $\cal W$ of closed walks in a graph $G$ satisfies the exclusive 3-walk property if for every two vertices $x$ and $y$ and every three $xy$-walks $W_1$, $W_2$ and $W_3$, we have:
$$\sigma_{\cal W}(W_1W_2^{-1})\cdot \sigma_{\cal W}(W_1W_3^{-1})\cdot\sigma_{\cal W}(W_2W_3^{-1})=+.$$
In other words an even number of the three closed walks $W_1W_2^{-1}$,
$W_1W_3^{-1}$, and $W_2W_3^{-1}$ is in $\cal W$.
\end{df}

Naserasr, Sopena, and Zaslavsky \cite{NSZ} present Theorem 9, which says that: 
A set $\cal W$ of closed walks in a graph $G$ is the set of negative closed walks in $(G,\sigma)$ for some choice of signature $\sigma$ if and only if
$\cal W$ satisfies the exclusive 3-walk property.
In their proof of Theorem 9, they also prove that if $\cal W$ satisfies exclusive 3-walk property then it is closed under rotation (Proposition 8[vi] in \cite{NSZ}).

In Section~\ref{C} we shall show that Theorem 9 in \cite{NSZ} is not valid and that it is possible that a set of closed walks satisfies exclusive 3-walk property and is not closed under rotation. In Section~\ref{N} we shall give a new characterization, namely:

\begin{thm}\label{NC}
A set $\cal W$ of closed walks in a graph $G$ is the set of negative closed walks in $(G,\sigma)$ for some choice of signature $\sigma$ if and only if
$\cal W$ is closed under rotation and satisfies the exclusive 3-walk property.
\end{thm}

\section{Counterexample}\label{C}
Consider a signed graph $(G,\sigma)$ with at least one negative cycle and a vertex $v_0\in V(G)$
which belongs to the negative cycle.
For example, let $(G,\sigma)$ be the cycle $(v_0,v_1,v_2)$ with one negative edge.

\begin{df}
Let us define three sets of closed walks in $G$:
\begin{itemize}
\item  ${\cal W}_\sigma$ - the set of all negative closed walks in $(G,\sigma)$.
\item ${\cal W}_{v_0}$ - the set of all closed walks which start at $v_0$.
\item ${\cal W}={\cal W}_\sigma - {\cal W}_{v_0}$.
\end{itemize}
\end{df}

\begin{lem}
\begin{itemize}
\item[(a)]  ${\cal W}_\sigma$ satisfies exclusive 3-walk property. 
\item[(b)] ${\cal W}$ satisfies exclusive 3-walk property. 
\item[(c)] ${\cal W}$ is not closed under rotation.
\end{itemize}
\end{lem}
\noindent {\bf Proof.} 

\noindent (b) If $x\ne v_0$, then it follows from (a). If $x=v_0$, then it is obvious. \hfill$\Box$

\begin{lem}
For no signature $\tau$, the family $\cal W$
is the set of negative closed walks in $(G,\tau)$. This implies that Theorem 9 in \cite{NSZ} is not valid.
\end{lem}
\noindent {\bf Proof.} The set ${\cal W}$ is not closed under rotation. On the other hand, for each signature $\tau$, the set of negative closed walks in $(G,\tau)$ is closed under rotation. \hfill $\Box$

Notice that also Proposition 8[v] in \cite{NSZ} is not valid.

\section{New characterization of negative closed walks}\label{N}

The proof of Theorem~\ref{NC} goes similarly as the proof of
Theorem 9 in \cite{NSZ}. However, the proof in \cite{NSZ} uses Proposition 8 which is not valid. For this reason one should use the following two lemmas instead.

\begin{lem}[see \cite{NSZ}, Proposition 8]\label{Prop7}
Let $G$ be a graph and let $\cal W$ be a set of closed walks which satisfies the
exclusive 3-walk property. Then $\cal W$ satisfies the following properties:
\begin{itemize}
\item[(i)] No trivial walk is in $\cal W$.
\item[(ii)] For any walk $W$, the closed walk $WW^{-1}$ is not in $\cal W$.
\item[(iii)] For any closed walk $W$, $\sigma_{\cal W}(W)=\sigma_{\cal W}(W^{-1})$.
\item[(iv)] For any pair of closed walks $W$ and $W'$ with the same starting point $v$, we have
 $\sigma_{\cal W}(WW')=\sigma_{\cal W}(W)\cdot\sigma_{\cal W}(W')$.
\end{itemize}
\end{lem}

\noindent{ Proof.}

\smallbreak
\noindent
(i) follows immediately from (ii).

\smallbreak
\noindent
(ii) Take $W_1=W_2=W_3=W$. By exclusive 3-walk property, $\sigma_{\cal W}(WW^{-1})=+$.

\smallbreak
\noindent
(iii) Let $v$ be the starting point of $W$. Take $x=y=v$, $W_2=W$, and let $W_1=W_3=e_v$, where $e_v$ is the trivial walk in $v$. Then $W_1W_2^{-1}=W^{-1}$, $W_1W_3^{-1}=e_v$, and $W_2W_3^{-1}=W$. Since
$\sigma_{\cal W}(e_v)=+$, we have $\sigma_{\cal W}(W)=\sigma_{\cal W}(W^{-1})$.

\smallbreak
\noindent
(iv) Take $x=y=v$, $W_1=W$, $W_2=W'^{-1}$, and $W_3=e_v$. Then $W_1W_2^{-1}=WW'$, $W_1W_3^{-1}=W$, and $W_2W_3^{-1}=W'^{-1}$. 
By exclusive 3-walk property:
$$\sigma_{\cal W}(WW')\cdot \sigma_{\cal W}(W)\cdot \sigma_{\cal W}(W'^{-1})
=\sigma_{\cal W}(WW')\cdot \sigma_{\cal W}(W)\cdot \sigma_{\cal W}(W')=+.$$
\hfill$\Box$

\begin{lem}[see \cite{NSZ}, Proposition 8(v)]\label{Prop8}
Suppose that $G$ is a graph and $\cal W$ is a set of closed walks which is closed under rotation and satisfies the exclusive 3-walk property.
Let $W$ be a closed walk starting at $y$ and $P$ be  an $xy$-walk.
Then $\sigma_{\cal W}(PWP^{-1})= \sigma_{\cal W}(W)$.
\end{lem}
\noindent{\bf Proof.} 
$PWP^{-1}$ is an rotation of $P^{-1}PW$, and by Lemma~\ref{Prop7}(ii) and (iv), 
$\sigma_{\cal W}(P^{-1}PW)= \sigma_{\cal W}(W)$. \hfill $\Box$

\end{document}